\documentclass[newstyle,twocolumn,proceedings,graphics]{rmaa}

\newcommand{\hii}{\ion{H}{2}}

\title{Ultracompact \hii{}  Regions}
\author{S. Kurtz and J. Franco
  \affil{Instituto de Astronom\'{\i}a, UNAM} }

\fulladdresses{
\item S. Kurtz: Instituto de Astronom\'\i a,
  UNAM-Morelia, Apdo. Postal 3-72, C.P. 58089, Morelia, Michoac\'an, M\'exico
  (skurtz@mailaps.org).
\item J. Franco: Instituto de Astronom\'\i a, UNAM, Apdo. Postal 70-264,
  C.P. 04510, M\'exico, D.F., M\'exico (pepe@astroscu.unam.mx).}

\shortauthor{Kurtz \& Franco}
\shorttitle{Ultracompact \hii{}  Regions}


\abstract{%

We review some recent observational results on the properties of 
ultracompact~\hii{} 
regions, in particular the presence of extended continuum emission surrounding
ultracompact sources and the discovery of a new class of so-called 
``Hypercompact'' \hii{} regions.
In addition, we discuss recent attempts to probe the density structure
within UC~\hii{} regions using the technique of spectral index analysis. }

\resumen{%
Hacemos una revisi\'on de los resultados observacionales recientes sobre
regiones \hii{} ultracompactas.  En particular, discutimos la emisi\'on 
extendida de algunas regiones \hii{} UC y el descubrimiento de una nueva
clase, denominada regiones \hii{} ``hipercompactas''.  Adem\'as,
describimos la obtenci\'on de la estructura de densidad usando la t\'ecnica
del an\'alisis del \'\i ndice espectral.

}

\listofauthors{S.~Kurtz \& J.~Franco}
\indexauthor{Kurtz, S.}
\indexauthor{Franco, J.}

\begin{document}
\maketitle

\section{Introduction}

\hii{}  regions are a relatively well-studied class of objects that are
usually classified as ultracompact, compact, and extended (or classical;
e.g., Habing \& Israel 1979). Ultracompact (UC)~\hii{} regions
have sizes of about $0.1$ pc, and are located in the inner,
high-pressure, parts of molecular clouds (see Kurtz et
al. 2000 for a recent review). Compact \hii{} regions have larger sizes,
0.1-0.3 pc, lower densities, and are located in more evolved zones of 
the clouds.  Extended \hii{} regions have sizes of up to several parsecs
and they represent the mature state of these objects. Giant and
supergiant \hii{} regions are observed in external galaxies, but they
represent a conglomeration of many individual \hii{} regions that have
already photoionized a large fraction of their parental giant
molecular clouds. We summarize the classes of \hii{} regions
in Table~1, and list the approximate physical parameters defining each
class (see Dyson \& Franco 2001 for a global review).  
The expansion of these objects is important for our
understanding of the evolution of one classification to another, 
as well as the structure and fate of clouds with active star formation. 

In the past three years there have been at least two quite interesting 
observational developments in the study of UC~\hii{} regions.
One deals with small-scale structures, the other with larger, 
more extended structures.  Both will challenge
theoretical models of UC~\hii{}  regions and may be fruitful areas for 
research in coming years.
In the following, we discuss these new results and their possible
implications for our understanding of the formation and evolution
of \hii{}  regions.

\section{Historical Perspective}

Compact \hii{}  regions were first identified as a class of objects by Mezger
et al.\  (1967), who described them as having sizes from 0.06 to 0.4~pc and 
electron densities close to 10$^4$~cm$^{-3}$.  
Qualitatively, they were described  as ``small, high-density \hii{}  regions 
{\it in extended \hii{}  regions of lower electron density''} (emphasis added).  
These ``extended'' regions had sizes of order 10~pc and densities of order
10$^2$~cm$^{-3}$.  

In the early 1990s about a half dozen VLA surveys were made of UC~\hii{} 
regions including work by Wood \& Churchwell (1989), Garay et al.
(1993), Kurtz, Churchwell \& Wood (1994), and Miralles, Rodr\'\i guez \& Scalise
(1994).  The Galactic Plane Surveys, summarized by Becker et al.
(1994) also contributed a large number of candidate UC~\hii{}  regions.
In total, several hundred UC~\hii{}  regions were identified.  These
surveys were typically made at wavelengths from 2 to 6~cm, in
configurations of the VLA that provided arcsecond resolution, and were
sensitive to structures up to 20-30$''$ in size.  This was a perfectly
reasonable observing strategy to use: people were looking for objects
expected to be $< 10''$ in size and with densities $n_e \sim
10^4$~cm$^{-3}$, which become optically thin at about 4~cm.
Nevertheless, two selection effects are evident: first, that only
structures smaller than 20-30$''$ are seen; second, that the highest
sensitivity is to sources with turnover frequencies at about 8~GHz
($\lambda$ = 4~cm).   Several groups have undertaken studies to
determine the significance of these selection effects;
the goal of this presentation is to overview their preliminary results.

\section{UC \hii{}  Regions with Extended Emission}

The primary studies of the effect of insensitivity to large-scale 
structures have been made by Kurtz et al. (1999) and by Kim \& Koo (1996, 
2001).
The former began with a random selection of 15 UC~\hii{}  regions and made
VLA observations sensitive to structures up to 3$'$ in size.  In 12 (80\%)
of these they found extended emission, and in eight (53\% of 15)  
they suggest (on morphological grounds) that the extended emission may have 
a direct physical relationship to the UC~\hii{}  region.  An example of such
extended emission is seen in the G35.20--1.74 region, shown in Figure~1.
In the latter study, Kim \& Koo chose 16 regions with high single-dish to
interferometric flux density ratios; they found extended emission in all 16
cases.  Although their work is not appropriate for a statistical study,
it is the largest dataset of UC~\hii{}  regions with extended emission.
Furthermore, they present radio recombination line (RRL) data that support
the idea that the extended emission is directly related to the ultracompact
emission.  This is an important point, because the physical relationship
between the extended and ultracompact emission has {\it not} been 
confirmed for the Kurtz et al. sample.

The presence of the extended continuum emission is not surprising.
Indeed, the original classification of compact \hii{}  regions outlined
in \S2 suggests that such emission is to be expected.  What has perhaps
not been fully appreciated until recently is that this extended emission
may be an essential part of UC~\hii{} regions, and must be accounted for
both when deriving physical parameters and applying theoretical models.
One possible scenario that would give rise to both ultracompact and 
extended ionized gas components is density structure within  molecular
clouds.  This idea was suggested by both Franco et al. (2000a) and by
Kim \& Koo (2001).  The latter provide a useful schematic of the idea,
which is shown in Figure~2.

\section{Hypercompact \hii{}  Regions}

The other selection effect, which gives preference to regions with 
emission peaks at $\sim 4$~cm, effectively limits existing \hii{} 
region surveys to objects with emission measures less than about 
10$^8$~pc~cm$^{-6}$.  This is because $\tau_\nu \propto \nu^{-2.1} EM$,
so that regions with emission measures greater than 
$4 \times 10^9$~pc~cm$^{-6}$ will remain optically thick into the 
millimeter regime.  Hence, centimeter flux densities will be significantly
lower than the peak flux density of the region.

Millimeter-wave observations with the VLA, using the recently developed
Q-band system, have identified a number of small, very high emission 
measure objects (e.g., G75.78+0.34-H$_2$O; see Figure~1 and Carral et al.
1997).  An inspection of Table~1 shows that these regions --- which we shall
refer to has ``hypercompact'' \hii{} regions --- are more than an order of
magnitude smaller and two orders of magnitude denser than UC~\hii{} regions.
This is substantially more pronounced than the difference between the compact
and ultracompact classifications, so we feel fully justified in defining a
new class of hypercompact (HC) regions.

Although a new taxonomic classification is clearly warranted, it is less
clear that a fundamentally new object has been discovered.  In particular,
it is unclear if HC~\hii{} regions are very young UC~\hii{} regions, that are
still confined to very small sizes and high densities, or if they are 
UC~\hii{} regions that formed in a particularly high density environment,
and which might be relatively old (see De~Pree, Rodr\'\i guez \& Goss 1995).
A possible clue lies in the remarkable overlap between candidate HC~\hii{}
regions (as indicated by very small sizes and optically thick centimeter
continuum spectra) and the relatively small group of broad RRL regions
(see Table~2).  This overlap suggests that HC~\hii{} regions may be in a
stage of rapid expansion, as would be expected for an extremely young \hii{}
region.  VLA observations are underway to test this hypothesis.

\section{Density Gradients in \hii{}  Regions}

For electron density distributions of power law form 
 $n_e \propto r^{-\omega}$, the spectral index $\alpha$ ($S_\nu \propto
\nu^\alpha$) depends on $\omega$ as $\alpha = (2\omega - 3.1)/(\omega - 0.5)$,
(Olnon, 1975).  Thus, multi-frequency radio observations provide a
means to probe the density structure of \hii{} regions.  

Franco et al. (2000b) use this technique to study three galactic
UC~\hii{} regions (see Figure~1) and report density gradients steeper
than $\omega$= 1.5.  G35.20--1.74, shown at the top of Figure~1, has
both compact and extended emission.  The original map (inset), made
with sub-arcsecond resolution, was sensitive only to structures
smaller than about 20$''$ (Kurtz et al. 1994) and has been convolved
with a 1\farcs2 $\times$ 0\farcs 9 Gaussian.  The rectangular boxes
indicate the integration areas used to obtain the spectral indicies
reported by Franco et al.  Subsequent lower resolution observations,
sensitive to structures up to 3$'$ in size, show the full extent of
the ionized gas in the region.  Spectra for the peak and tail regions
are shown in Figure~3.  The need for a non-uniform density model is
evident from the peak spectrum, which is reproduced in Figure~4, along
with theoretical spectra corresponding to a gaussian density distribution 
and to a uniform distribution.  The latter two dramatically under-estimate 
the observed flux densities.

The G9.62+0.19-E region, shown at lower left in Figure~1 in a map adapted
from Testi et al. (2000), has a spectral index corresponding to a density 
gradient of $n_e \propto r^{-2.5}$.  The spectrum of this region is shown
in Figure~5, where it will also be noted that it is a candidate HC~\hii{} 
region.  The spectrum appears to be consistent with free-free emission
that remains optically thick to 2.7~mm.  Alternatively, the free-free
emission may be turning over at the 7~mm point, while the 2.7~mm point
reflects thermal emission from warm dust.

G75.78+0.34-H$_2$O, shown at lower right in Figure~1 in a map adapted from
Carral et al. (1997), has a density gradient exponent of --4, based on
the spectral index analysis of Franco et al.  They suggest that a 
gaussian density distribution or the contribution of dust emission at
high frequencies may cause this large (and probably incorrect) value.

If no mechanism acts to maintain density inhomogeneities within the \hii{} 
region, they will be smoothed
out on the order of a sound-crossing time
(Rodr{\'\i}guez-Gaspar, Tenorio-Tagle \& Franco 1995).
This is also true for the initial density gradients, which are smoothed
over time by the expansion. Thus, the density gradients that Franco et al. 
report are lower limits to the original distribution.

\begin{table*}\centering
\caption[]{Physical Parameters of \hii{}  Regions}
\begin{tabular}{rrrrr}
\hline
Class of        & \multicolumn{1}{c}{Size} & \multicolumn{1}{c}{Density} & 
\multicolumn{1}{c}{EM} & Ionized Mass \\
Region   &  \multicolumn{1}{c}{(pc)} & 
\multicolumn{1}{c}{(cm$^{-3}$)} & 
\multicolumn{1}{c}{(pc cm$^{-6}$)} & 
\multicolumn{1}{c}{(M$_\odot$)}\\
\hline
Hypercompact &  $\sim0.003$ &  $\ga 10^6$ &  $\ga 10^{10}$& $\sim 10^{-3}$ 
\\
Ultracompact    &  $\la 0.1$   &  $\ga 10^4$  & $\ga 10^7$ &$\sim 10^{-2}$ \\
Compact  &  $\la 0.5 $  &  $\ga 5 \times 10^3$  &   $\ga 10^7$ & $\sim 1$ \\
Classical & $\sim 10$ & $\sim 100$& $\sim 10^2$ & $\sim 10^5$\\
Giant & $\sim 100$ &  $\sim 30$ & $\sim 5 \times 10^5$ & $10^3$--$10^6$\\
Supergiant & $>$100 & $\sim 10$ & $\sim 10^5$  & $10^6$--$10^8$  \\            
              \hline
\end{tabular}
\end{table*}

\begin{figure*}
  \includegraphics[width=\textwidth]{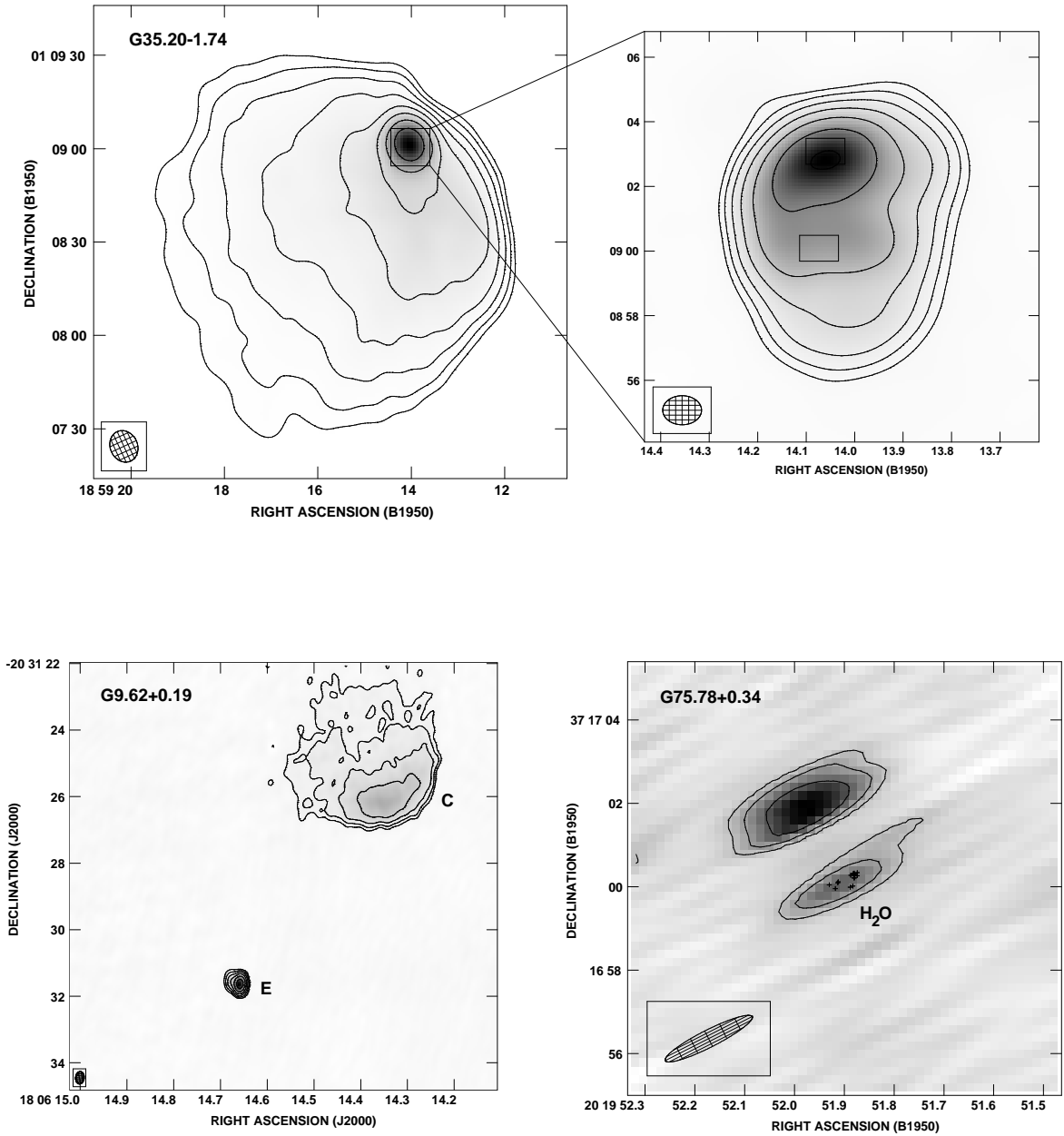}
  \caption{ This figure summarizes the essential recent developments
    in the study of UC~\hii{}  regions.  Above is shown the G35.20--1.74
    UC~\hii{}  region, which has a cometary morphology at at both arcsecond
    scales (top right) and at arcminute scales (top left).  In addition,
    the region shows evidence for density gradients, based on a spectral
    index analysis of the flux densities (see figures 3 and 4).
    At bottom, the regions G9.62+0.19-E and G75.78+0.34-H$_2$O are both 
    candidate hypercompact \hii{}  regions and both show evidence for
    density gradients.}
\end{figure*}

\begin{figure*}
    \includegraphics[width=\textwidth]{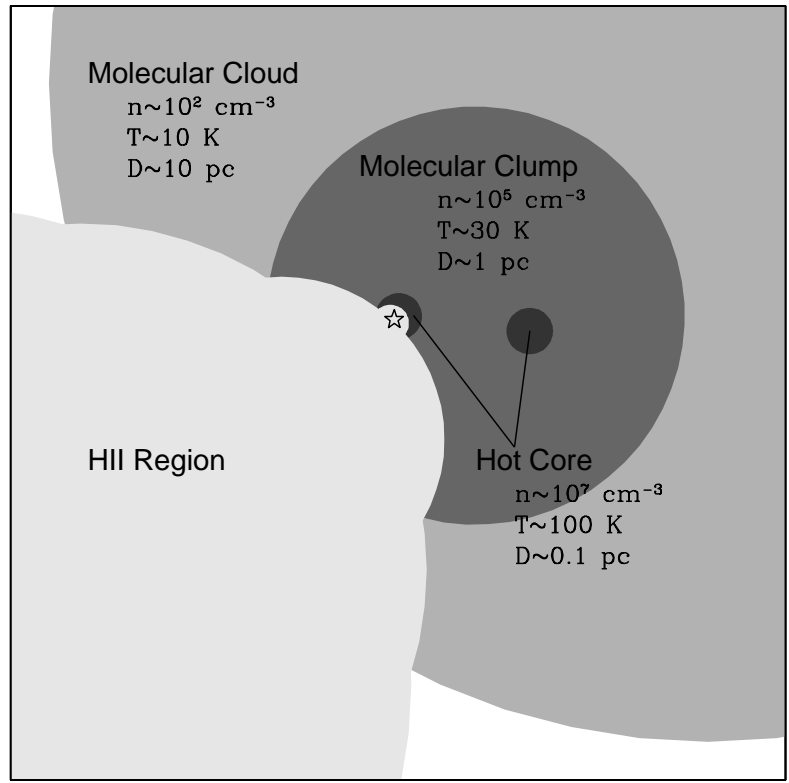}
    \caption{Schematic representation of the model proposed by Kim \& Koo
     to explain the extended emission around UC~\hii{} regions. The figure is not 
     to scale; taken from Kim \& Koo (2001), their Figure~8.   
     Franco et al. (2000a) also suggest density structure within the
      parent molecular cloud as a possible cause of the extended emission.}
\end{figure*}

\begin{figure}
\includegraphics[angle=0,width=\linewidth]{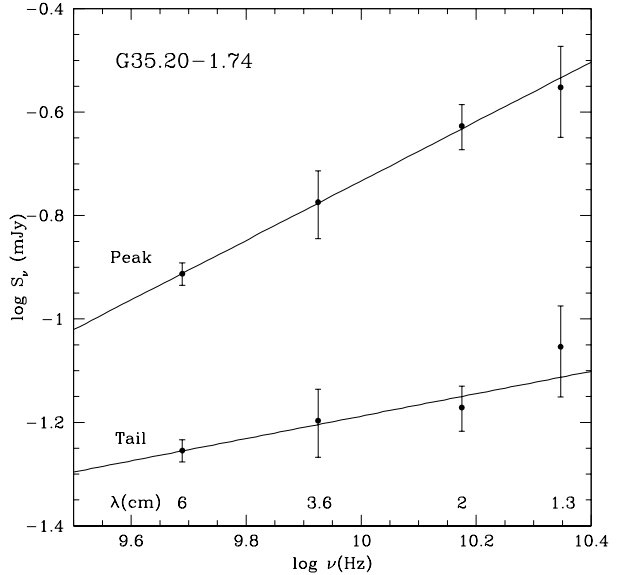}
\caption{ The flux density distribution of G35.20$-$1.74.
        Flux densities are plotted for 4.885, 8.415, 14.965, and 22.232~GHz.
        The solid lines are least-squares fits, giving a spectral index
        at the peak position of $\alpha=0.6\pm0.1$ and at the tail position
        of $\alpha=0.2\pm0.1$. The lower spectral index in the tail region
        may be indicative of a transition from a steeper core gradient to
        a uniform inter-clump medium.}
\end{figure}

\begin{figure}
\includegraphics[angle=0,width=\linewidth]{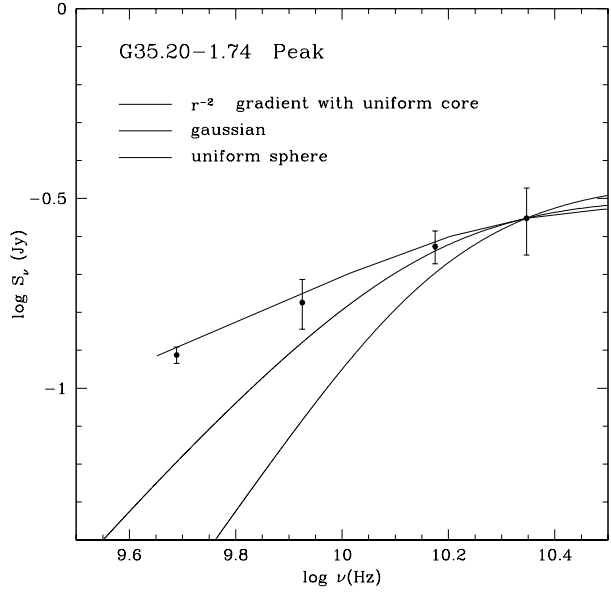}
\caption{Theoretical spectra for G35.20$-$1.74.
        The data points are for the peak position, as shown in Fig.~1a.
        The three curves correspond to the expected flux densities for a
        uniform sphere of ionized gas, for a gaussian distribution, and for
        a flattened power law (i.e., a gradient with a uniform core).  In
        accordance with the spectral index of the peak position, we adopt a
        density gradient of $r^{-2}$.  All three curves have been scaled to
        match the flux density at 1.3~cm, with the assumption that this point
        corresponds to the turn-over frequency.}
\end{figure}

\begin{figure}
\includegraphics[angle=0,width=\linewidth]{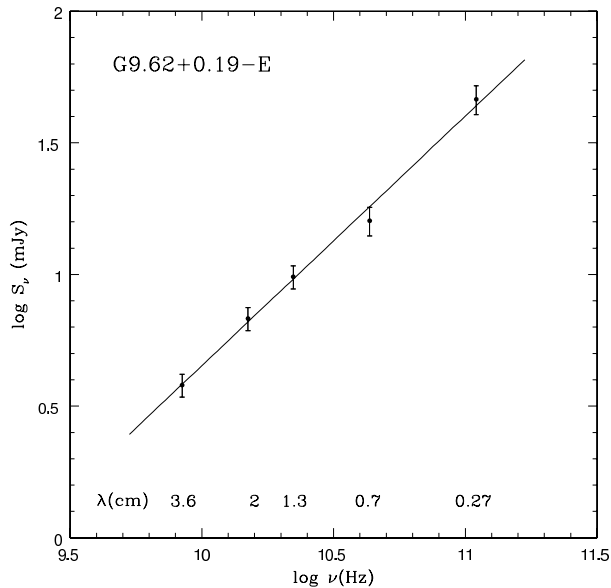}
\caption{The flux density distribution for G9.62+0.19-E.
        Flux densities are plotted for 8.415, 14.965, 22.232, 43.34, and 
        110~GHz.
        The solid line is a least-squares fit, yielding a spectral index of 
        $\alpha=0.95\pm0.06$, which suggests a density gradient of
        $n_e \propto r^{-2.5}$.}
\end{figure}

\begin{table}
  \setlength{\tabnotewidth}{0.8\columnwidth} 
  \tablecols{3} 
  \begin{center}
    \caption{Known Broadline Regions\tabnotemark{a}}  
    \begin{tabular}{lccc}\hline\hline
             &  FWHM            & Spectral    \\
Source       &  (km~s$^{-1}$)   & Index  & Ref  \\ \hline
NGC 7538   & 180 & ... & 1     \\
G25.5+0.2  & 161 & ... & 2    \\
Sgr B2     & 80  & 0.95 & 3    \\
W49 AA     & 50  & 0.6  & 4    \\
W49 AB     & 60  & 1.1  & 4    \\
W49 AG     & 45  & 2    & 4    \\
M17-UC1    & 47  & 1.1  & 5   \\
      \hline\hline
      \tabnotetext{a}{Table is adapted from Johnson et al. (1998).}
      \tabnotetext{[1]}{Gaume et al. 1995.}
      \tabnotetext{[2]}{Shepherd et al. 1995.}
      \tabnotetext{[3]}{De Pree et al. 1996.}
      \tabnotetext{[4]}{De Pree et al. 1997.}
      \tabnotetext{[5]}{Johnson et al. 1998.}
    \end{tabular}
  \end{center}
\end{table}

\acknowledgements We are very grateful to P. Hofner and G. Garc\'\i a-Segura
for their contributions to the work discussed in this paper. 
We also acknowledge useful discussions with N. Mohan, S. J. Arthur, and A. Esquivel.
Financial support for this research has been provided by DGAPA-UNAM 
(project number IN117799) and by CONACyT, Mexico.


\end{document}